\documentclass[aps,prb,twocolumn,groupedaddress]{revtex4}

\usepackage{overpic}
\usepackage{color}
\usepackage{amsmath}
\usepackage{bm}
\usepackage{mathrsfs}

\usepackage{overpic}
\usepackage{mathrsfs}

\usepackage{subfig}
\usepackage{multirow}

\usepackage{afterpage}
\usepackage{float} 

\DeclareMathAlphabet{\mathscrbf}{OMS}{mdugm}{b}{n}

\begin{document} 

\title{Importance of relativistic effects in electronic structure and \\
thermopower calculations for Mg$_2$Si, Mg$_2$Ge and Mg$_2$Sn.}



\author{K. Kutorasinski}
\email[]{kamil.kutorasinski@fis.agh.edu.pl}
\author{B. Wiendlocha}
\author{J. Tobola}
\author{S. Kaprzyk}
\affiliation{AGH University of Science and Technology,\\ Faculty of Physics and Applied Computer Science,
Al.~A.~Mickiewicza 30, 30-059 Krakow, Poland}


\date{\today}

\begin{abstract}

We present a theoretical study of the influence of the relativistic effects 
on electronic band structure and thermopower of Mg$_2X$ ($X=$ Si, Ge, Sn) 
semiconductors. The full potential Korringa-Kohn-Rostoker (KKR) method is used, and 
the detailed comparison between the fully relativistic and semi-relativistic 
electronic structure features is done. We show that the spin-orbit (S-O) 
interaction splits the valence band structure at $\Gamma$ point in good 
agreement with the experimental data, and this effect strongly depends on $X$ atom. 
The S-O modifications of the topology of the $\Gamma-$centered hole-like 
Fermi surface pockets lead to a change in electron transport properties, 
which are investigated using the Boltzmann approach. 
In addition, the simple and efficient method is presented for the calculation of density of states effective mass $m^*$, 
and then used to examine the impact of relativistic effects on $m^*$. It is found 
that S-O coupling of the valence bands reduces effective mass and therefore significantly 
lowers the thermopower, primarily in Mg$_2$Sn, but also in Mg$_2$Ge. 
A detrimental influence of the S-O interaction on thermoelectric performance 
of $p-$type Mg$_2X$ is analyzed in function of temperature ($10-900$ K) and 
carrier concentration ($10^{18}-10^{22}$ cm$^{-3}$).
Interestingly, similar calculations in $n-$type Mg$_2$X, show negligible 
effect of the S-O interaction on lowest conduction bands and 
consequently also on the Seebeck coefficient.


\end{abstract}


\maketitle

\section{Introduction}

\label{sec:Introduction}

The well-known Mg$_2X$ ($X=$ Si, Ge, Sn) thermoelectric compounds 
\cite{winkler55,morris58} and their alloys \cite{nikitin61,labotz63,noda92} have 
attracted renewed attention~\cite{zaitsev06,tani05} as they are composed 
of abundant, low-cost (except germanium) and relatively 
non-toxic elements. These compounds exhibit large Seebeck coefficient ($S$), 
quite high electrical conductivity ($\sigma$) and low thermal 
conductivity ($\kappa$), yielding high efficiency of thermoelectric (TE) 
conversion, which is commonly captured in the dimensionless figure of merit 
$zT = \frac{\sigma S^2}{\kappa} T$ ($T$ is the absolute  temperature).
The fact that some of these materials have $zT > 1$ in the mid-temperature 
range (400-800 K) and the lowest density amongst all efficient 
thermoelectrics~\cite{samunin13}, may prove to be decisive for 
ground transportation and space applications of these materials 
\cite{fml13}.

In particular, $n-$type Mg$_2$Si$_{1-x}$Sn$_x$ solid solutions have been found 
to be the most favorable in terms of TE energy conversion as they have the 
lowest thermal conductivity due to the maximum mass difference between 
its components and quite large effective mass of carriers \cite{zaitsev06}.
It was reported that more complex quasi-quaternary solid solutions 
Mg$_2$Si$_{1-x-y}$Sn$_x$Ge$_y$ exhibited even better TE performance 
upon $n-$type doping ($zT \sim 1.4$)\cite{khan13}. Conversely, $p-$type 
Mg$_2$Si-based materials show markedly poorer TE performance 
($zT<0.4$) and there are only few impurities 
(Ga\cite{mouko11} or Ag\cite{tani05,mars09}) which allow to turn 
these systems to the hole-like electrical conductivity and positive thermopower.

In order to elucidate unusual electron transport properties of Mg$_2$(Si-Ge-Sn) 
thermoelectrics, {\it ab initio} electronic structure calculations have been recently 
reported\cite{liu12,singh12,tan12,kk13}. They mostly focused on so-called 
convergence of two conduction bands \cite{tobola10,liu12,kk13}, which appeared near 
X point in the Brillouin zone of Mg$_2$Si$_{1-x}$Sn$_x$ for $x \sim 0.7$. 
Indeed, this electronic structure feature, expected to be responsible for overall enhancement 
of TE properties of these compounds, was supported by computations of 
relevant transport coefficients within the Boltzmann transport theory\cite{singh12,kk13}, 
but neglecting relativistic effects. We will show that the S-O interaction is crucial 
to reliably interpret especially $p-$type materials. Other theoretical works attempted to 
search for efficient $n-$ and $p-$ dopants to allow tuning and optimizing TE properties 
of these materials \cite{zwolenski11,fml13}.

This work presents results of the first principles calculations performed 
by fully relativistic and semi relativistic Korringa-Kohn-Rostoker (KKR) 
methods combined with the Boltzmann transport approach to calculate 
thermopower in function of temperature and carrier concentration. 
We show that the S-O interaction strongly affects top valence bands 
near $\Gamma$ point (best seen in Mg$_2$Sn) and the calculated values 
of the S-O splitting remain in good agreement with available 
experimental data. On the whole, the S-O coupling has a detrimental effect 
on $p-$type thermopower even at elevated temperature (300-400 K) and near optimal 
carrier concentration ($\sim 10^{20}$ cm$^{-3}$). On the other hand, the influence 
of the S-O coupling on conduction bands in $n-$type Mg$_2X$ 
is negligible whatever temperature and carrier concentration.

The paper is organized as follows. Key formulas connecting 
the electronic band structure obtained from Korringa-Kohn-Rostoker (KKR) method 
and the electron transport coefficients studied within the Boltzmann 
approach, as well as computational details are given 
in Sec.~\ref{sec:details}. Sec.~\ref{sec:results} presents our results 
containing mainly (i) an analysis of the hole-like and electron-like Fermi surface 
shapes, (ii) a determination of effective mass of carriers using a
simple and efficient procedure, (iii) a discussion of Seebeck coefficient 
vs. temperature and concentration. All results derived from semi-relativistic 
and fully relativistic KKR calculations are compared in systematic way.
The paper is concluded in Sec.~\ref{sec:Conclusions}.

\section{Theoretical Details}
\label{sec:details}

\subsection{KKR band structure}

The full potential KKR\cite{Kohn54,Butler76,Kaprzyk82,Kaprzyk90,Bansil99,Stopa04} 
method based on the Green's function multiple scattering theory was implemented to calculate electronic band structure. 
The reduction of the Dirac equation to the so-called semi-relativistic (SR) version was done with adopting technique developed by 
Koelling and Harmon\cite{KOELLING} and extended by Ebert\cite{Ebert}. With this procedure the resulting set of coupled 
radial equation after dropping the spin-orbit interaction are similar to the non-relativistic one, but retains all other 
kinematic effects such as mass-velocity, Darwin contribution and higher order terms. 
The full relativistic (FR) method was done by directly solving the four component Dirac equation without any simplification. 
This allows to calculate band structure with or without S-O interaction, showing its influence directly. Electronic structure 
calculations were performed with previously presented details\cite{Bansil99,Stopa06}.

As already discussed in literature~\cite{singh12,tran09}, LDA (and GGA as well) tends to underestimate band gaps also in Mg$_2X$ 
compounds. In recent works\cite{singh12,fml13} it was shown that the application of the modified Becke-Johnson semi-local exchange 
potential of Tran and Blaha \cite{tran09} yielded the gap values close to the experimental ones but without important changes 
in shape of bands. Thus, in this work, the standard LDA (with the Perdew and Wang\cite{PW92} formula for exchange-correlation potential) 
is employed in KKR calculations and the energy gaps are expanded to experimental values\cite{Handbook} (see, Table~\ref{table:tab1_1}) to 
allow for reasonable discussion of transport properties in function of temperature. 

\subsection{Electron transport}

The Boltzmann transport theory, which can successfully connect atomic level properties of materials with the macroscopic transport 
coefficients\cite{Thonhauser,Madsen,Chaput,kk13} is used for calculation of thermopower. Within this approach\cite{Ashcroft} the 
Seebeck coefficient can be expressed as

\begin{equation}\label{eq:s}
S=-\dfrac{1}{eT}\dfrac{\mathscr{L}^{(1)}}{\mathscr{L}^{(0)}},
\end{equation}
where
\begin{equation}
\mathscr{L^{(\alpha)}}=\int{d\mathscr{E}\left(-\dfrac{\partial f}{\partial \mathscr{E}}\right)(\mathscr{E}-\mu_c)^\alpha}\sigma(\mathscr{E}).
\label{eq:S}
\end{equation}
$\sigma(\mathscr{E})$ is an energy-dependent conductivity, commonly called the transport function (TF). 
Chemical potential $\mu_c=\mu_c(T,n_d)$ depends on temperature $(T)$ and doping ($n_d$), where an extra carrier concentration $n_d$ can be 
positive ($n-$type doping, e.g. Sb or Bi in Mg$_2X$) or negative ($p-$type doping, e.g. Ga in some Mg$_2$(Si-Ge) alloys\cite{mouko11}). 
In the present work, the rigid band model\cite{RBM} is used to mimic $n-$ and $p-$type behaviors, which allows to focus on the S-O 
effect on charge carrier transport in case of electron-like and hole-like doping, respectively. 

The transport function, within the relaxation time approximation, has the form of the $\mathbf{k}$-space integral over $n$ electronic 
bands: \cite{Ashcroft}
\begin{equation}
\sigma(\mathscr{E})=e^2\frac{1}{3}\sum_n\int\frac{d \mathbf{k}}{4 \pi^3}\tau_n(\mathbf{k})\mathbf{v}_n(\mathbf{k})^2\delta(\mathscr{E}-\mathscr{E}_n(\mathbf{k})),
\label{eq:sigmaall}
\end{equation}
where $\tau$ is the electron relaxation time and $\mathbf{v}_n(\mathbf{k}) =\hbar^{-1}\nabla_k\mathscr{E}_n(\mathbf{k})$ denotes the electron velocity.
Representing the band structure as isoenergetic surfaces\cite{Lorensen} ($\mathscr{E}_n(\mathbf{k})\rightarrow S_n(\mathscr{E})$) and 
employing the commonly used constant relaxation time approximation ($\tau_n(\mathbf{k})=\tau_0$) allow to convert TF into the form more convenient 
for numerical computation:
\begin{equation}\label{eq:sig}
\sigma_{\tau}(\mathscr{E})=\tau_0\frac{e^2}{\hbar}\frac{1}{3}\sum_n\int\limits_{S_n(\mathscr{E})}\frac{dS}{4\pi^3}|\mathbf{v}(S_n(\mathscr{E}))|.
\end{equation}
It is worth mentioning that upon substituting Eq.~\ref{eq:sig} into Eqs.~\ref{eq:sigmaall} and \ref{eq:S}, $\tau_0$ cancels out so that
thermopower is independent of $\tau_0$ in the constant relaxation time
approximation.

In a similar way, the density of states (DOS) function can be calculated  
\begin{equation}
g(\mathscr{E})=\sum_n\int\limits_{S_n(\mathscr{E})}\frac{dS}{4\pi^3}\frac{1}{|\nabla_k\mathscr{E}_n(\mathbf{k})|},
\end{equation}
and such representation enables to decompose the DOS (and also TF) into contributions from each $n$-th band.

To facilitate the discussion of the Seebeck coefficient, the effect of the S-O interaction on the effective mass is investigated. 
Accordingly, we propose a simple and efficient way of calculating the energy dependent effective mass. DOS effective mass is defined for the 
parabolic band $\mathscr{E}(\mathbf{k})=\hbar^2\mathbf k^2/2m$ through the formula:
\begin{equation}
g(\mathscr{E})=\frac{m}{\hbar^2\pi^2}\sqrt{\frac{2m\mathscr{E}}{\hbar^2}}.
\end{equation}
To generalize this concept to any other, non-parabolic case, the effective mass becomes energy (or carrier concentration) dependent $m=m(\mathscr{E})$. 
The effective mass can be extracted from the DOS formula and combined with its energy derivative
\begin{equation}
\label{eq:mass}
m(\mathscr{E})=m_em^*(\mathscr{E})=\hbar^2\sqrt[3]{\pi^4g(\mathscr{E})g'(\mathscr{E})},
\end{equation}
where $m_e$ denotes free electron mass.

In this way, the dimensionless DOS effective mass ($m^*$) can be found at every energy point, at which DOS was calculated. 
This approach can also be seen as fitting the effective mass with the parabolic band at every $\mathscr{E}$ and $\mathbf{k}$-points 
separately and then averaging over isoenergy surface $S_n(\mathscr{E})$. Practically, the computation of $m^*$ using this scheme requires highly accurate DOS function to avoid problems 
with numerical noise appearing when differentiating. Here, DOS is calculated with $\Delta\mathscr{E}\sim9$~meV energy resolution, 
and $m^*(\mathscr{E})$ resulting from Eq.~\ref{eq:mass} in final step is smoothed by convolution with the Gauss function with standard deviation $3\times\Delta\mathscr{E}$. 
Comparison of the raw and smoothed results is done in Fig.~\ref{fig:meff}. 

\begin{figure*}[htb]
\includegraphics[width=0.98\textwidth]{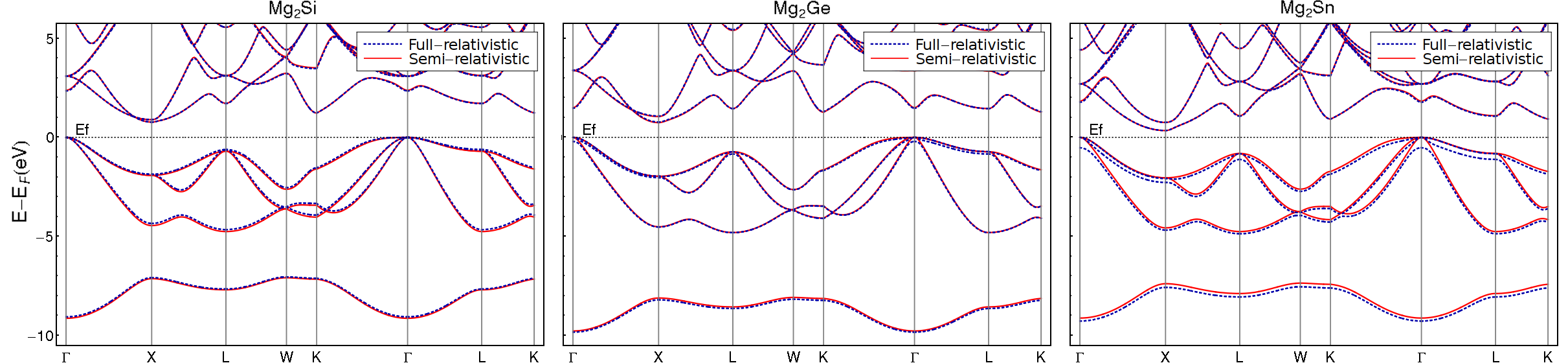}
\caption{(Color online) Electron dispersion curves for Mg$_2$Si (left), Mg$_2$Ge (middle) and Mg$_2$Sn (right) as resulted from semi-relativistic and 
full-relativistic KKR calculations. In each figure, the Fermi level $E_F$ was shifted to top of valence bands 
($\Gamma$ point) and energy gaps were expanded to the experimental values (see text and Table~\ref{table:tab1_1}).}
\label{fig:Bands1}
\end{figure*}

\section{ Results }
\label{sec:results}
\subsection{Band Structure}

\begin{figure}[b]
\centering
\includegraphics[width=0.468\textwidth,trim=0 0 0 0]{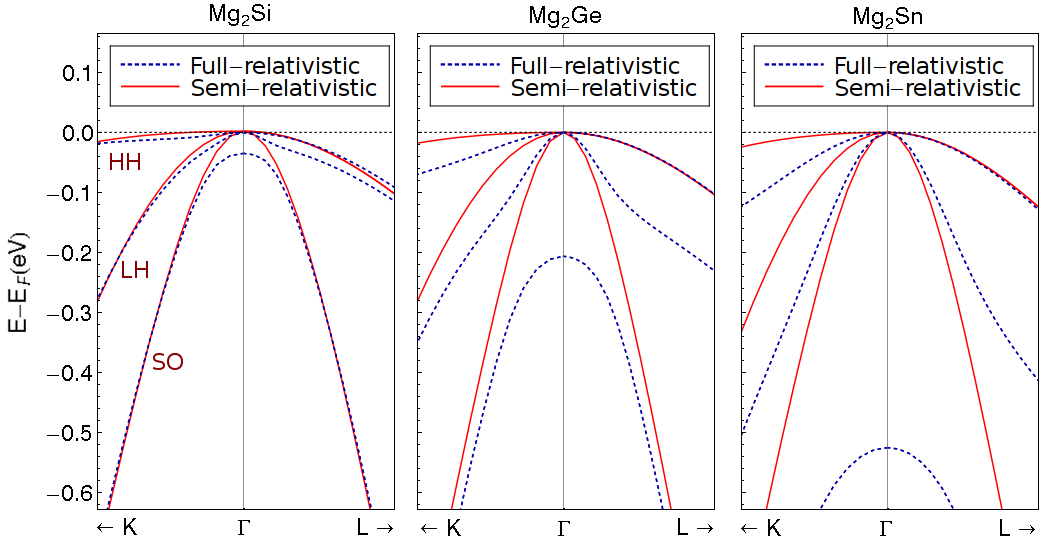}
\caption{(Color online) Zoom of the three valence bands of Mg$_2X$ along the K-$\Gamma$-L direction as computed from SR and FR calculations. 
The spin-orbit interaction removes the degeneracy of bands at $\Gamma$, and the largest splitting is observed for the Sn case. 
Subtle topological effects are evidently detected for the LH band (see text).}
\label{fig:Bands4}
\end{figure}

\begin{figure*}[hbt]
\centering
\begin{tabular}{lcr}
\subfloat[p-type Mg$_2$Sn]{%
\includegraphics[width=0.23\textwidth]{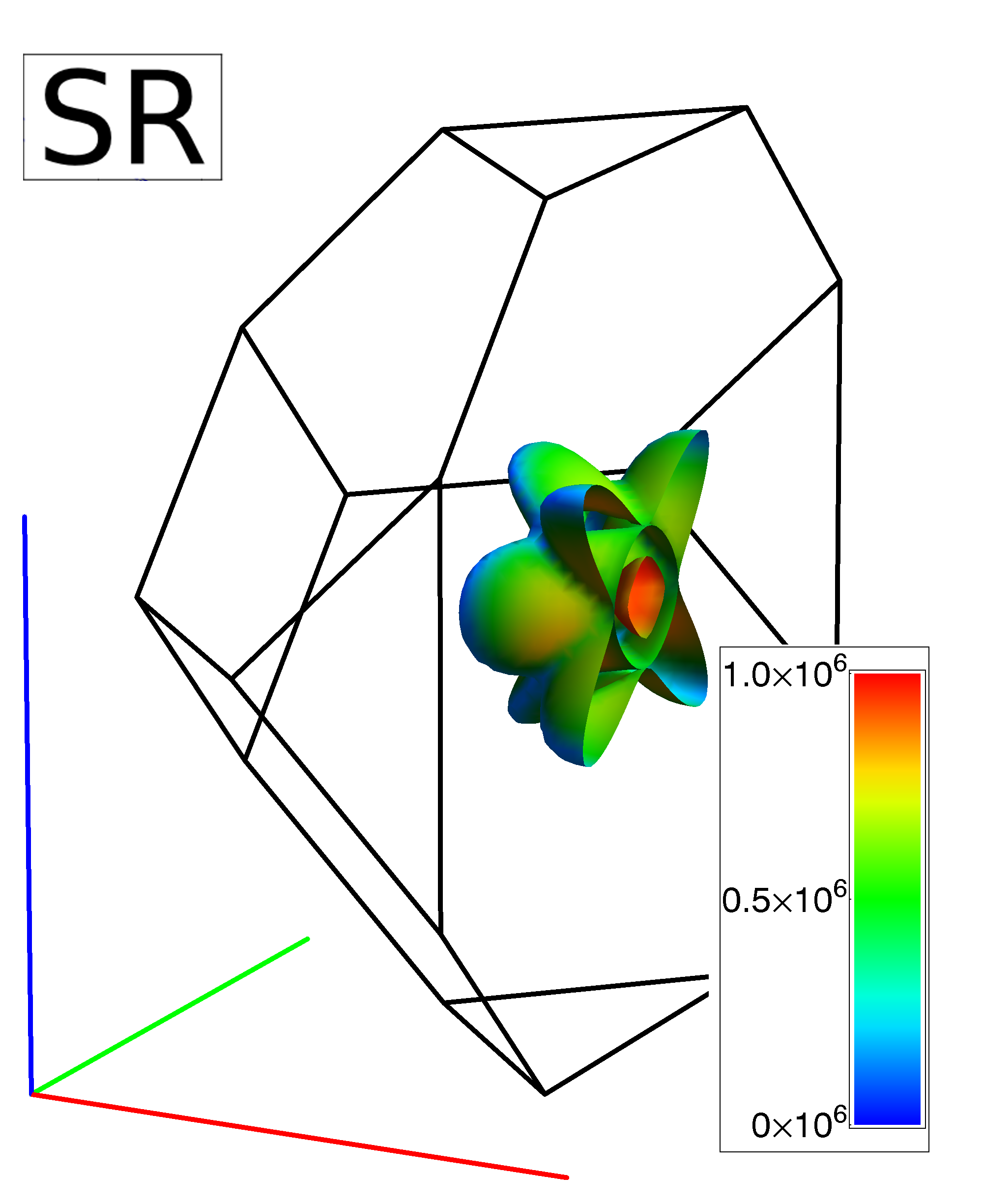}
\includegraphics[width=0.23\textwidth]{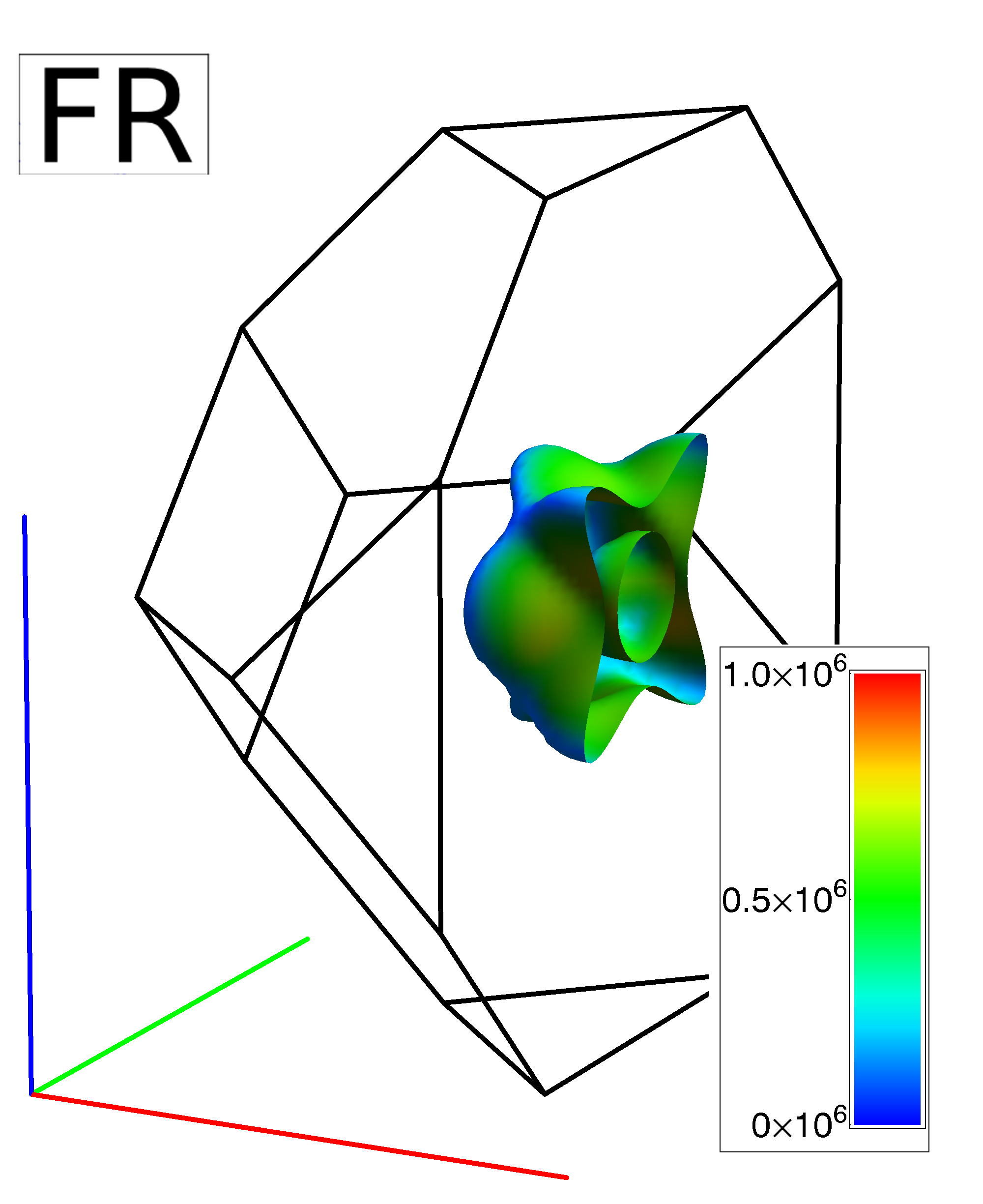}
}& \ \ \ \ &
\subfloat[n-type Mg$_2$Sn]{%
\includegraphics[width=0.23\textwidth]{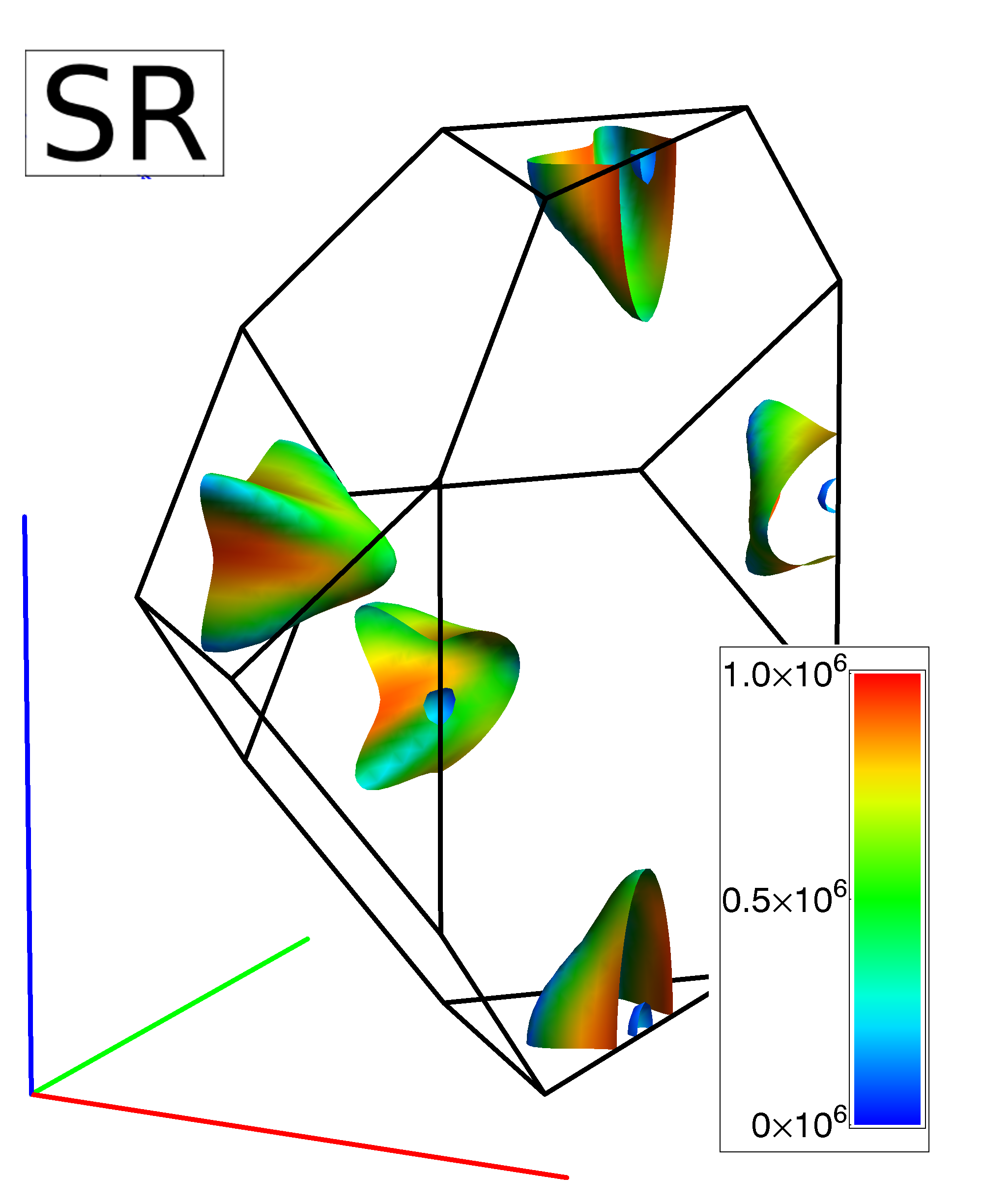}
\includegraphics[width=0.23\textwidth]{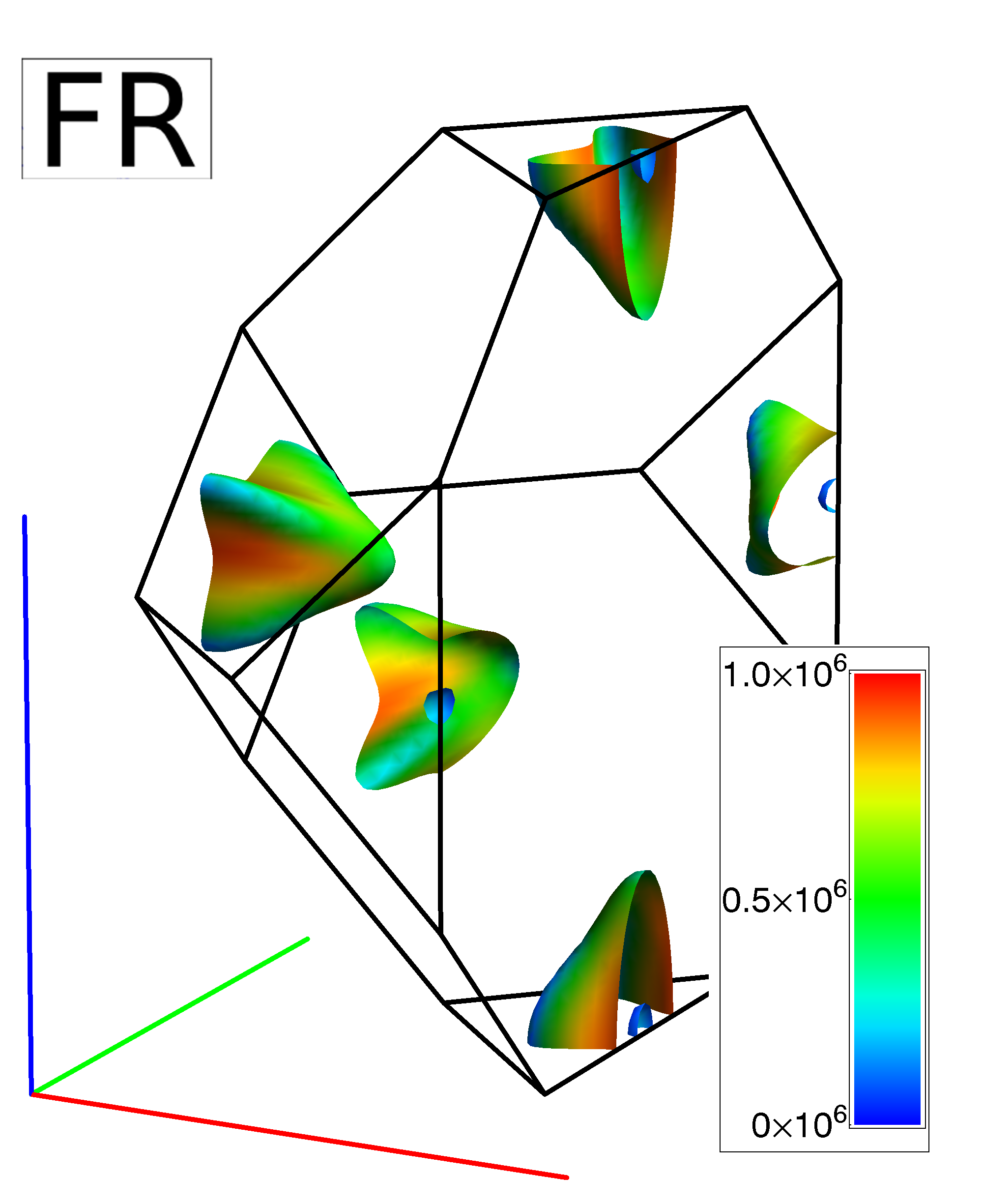}
 }
\end{tabular}
\caption{(Color online) Fermi surfaces of $p-$type (a) and $n-$type (b) Mg$_2$Sn plotted for carrier 
concentration $10^{21}$~cm$^{-3}$. In both cases semi-relativistic (SR) and fully relativistic (FR) results 
are compared. Electron velocities (in m/s) are represented by colors.}
\label{fig:FS}
\end{figure*}



\begin{table}[t]
\caption
{The energy band gap in Mg$_2X$ compounds calculated within LDA (E$^{LDA}_g$) from full-relativistic (FR) or semi-relativistic (SR) KKR methods and compared 
to experimental values (E$^{exp}_g$)~\cite{Handbook}. Experimental lattice constants\cite{zaitsev06} $(a)$ are also given.}
\centering
\begin{tabular}{lcccc}
\hline\hline
\multicolumn{1}{c}{}&\multicolumn{1}{c}{\multirow{2}{*}{$a$ (\AA) \ \ \ \ }}& \multicolumn{2}{c}{E$^{LDA}_g$(eV)}&\multirow{2}{*}{ \ \ \ \ E$^{exp}_g$ (eV)}\\\cline{3-4}
  & \ \ \ \  \ \ \ \ &\ \  \ \ SR \ \ \ & \ \ \ FR \ \ \ & \ \   \\
\hline
Mg$_2$Si  \ \ \ \ & 6.336 \ \ \ \ & 0.32  & 0.33 & 0.78 \\
Mg$_2$Ge & 6.385 \ \ \ \ & 0.21  & 0.23 & 0.72\\
Mg$_2$Sn & 6.765 \ \ \ \ & -0.17  & -0.25 & 0.35\\ [.3ex]
\hline\hline
\end{tabular}
\label{table:tab1_1}
\end{table}

\begin{table}[b]
\caption{The values of spin-orbit splitting at $\Gamma$ point in Mg$_2X$ compounds.}
\centering
\begin{tabular}{lccc}
\hline\hline
  & \ \ \ \  Mg$_2$Si \ \ \ \ & \ \ \ \ Mg$_2$Ge \ \ \ \ &  \ \ \ \ Mg$_2$Sn \ \ \ \ \\
\hline
Calculated (meV) \ & 36  & 208  & 525  \\
Measured (meV) \ & 30\footnotemark[1]  & 200\footnotemark[1]  & 480\footnotemark[1],600\footnotemark[2]  \\ [.3ex]
\hline\hline
\end{tabular}
\footnotetext[1]{Ref.~\onlinecite{vazquez1968}}
\footnotetext[2]{Ref.~\onlinecite{Lott1965}}
\label{table:tab1}
\end{table}

Electronic band structures of Mg$_2X$ as calculated from the fully relativistic (FR) and semi-relativistic (SR) KKR methods, 
are shown in Fig.~\ref{fig:Bands1}. Generally, electronic structure below the energy gap consists of four occupied bands. 
The lowest lying and separate band, essentially of $s-$symmetry, is located well below the Fermi energy and other three 
bands are forming main block of valence states (strongly hybridized $s-$Mg and $p-X$ states), with the bands maxima 
at $\Gamma$ point in Brillouin zone (BZ). These three bands are labeled as heavy hole (HH), light hole (LH) and split-off (SO) bands, 
respectively. 

The spin-orbit coupling removes the band degeneracy at $\Gamma$ point, pushing the SO band down in energy in different 
way in Mg$_2X$ compounds.
The conduction bands have the lowest energy value at the X point in BZ, yielding an indirect band gap in all Mg$_2X$ compounds. 
It was shown~\cite{kk13,zaitsev06} that relative position of the two lowest conduction bands 
mutually change with $X$ atom in Mg$_2X$, which is clearly seen in the Fig.~\ref{fig:Bands1}.

\begin{figure*}[htb]
\includegraphics[width=0.45\textwidth]{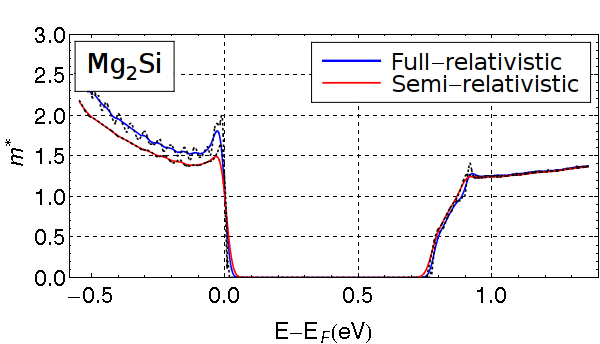} \ \ 
\includegraphics[width=0.45\textwidth]{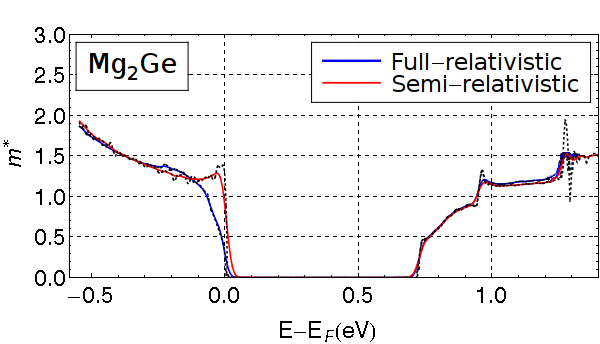}\\
\includegraphics[width=0.45\textwidth]{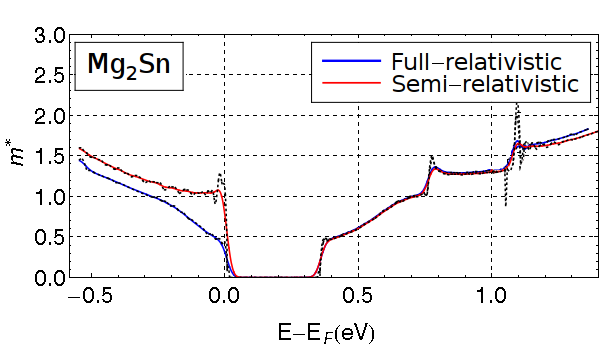} \ \
\includegraphics[width=0.22\textwidth]{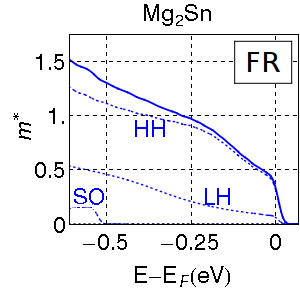}
\includegraphics[width=0.22\textwidth]{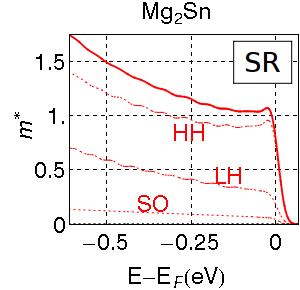}
\caption{(Color online) DOS effective mass of carriers near the band gap in Mg$_2X$ compounds. Bottom-right panel shows the effective hole mass corresponding to the heavy-hole (HH) 
light-hole (LH) and split-off (SO) bands for full-relativistic (FR) and semi-relativistic (SR) results. The spin-orbit interactions mostly affect mass of heavy holes.}
\label{fig:meff}
\end{figure*}

As expected, the importance of relativistic effects gradually increases with increasing atomic number of the $X$ element. 
In Fig.~\ref{fig:Bands1} the difference between SR and FR, as induced by the S-O coupling, can be easily detected for Mg$_2$Sn  
and also Mg$_2$Ge, while it is hardly visible in this energy range for Mg$_2$Si. The S-O interaction manifests most strongly for 
the valence bands around the $\Gamma$ point. It is best seen in Fig.~\ref{fig:Bands4} where tentatively three relativistic effects 
can be observed.

The first effect (as above-mentioned) is removing the degeneracy of electronic states at $\Gamma$ point by the S-O splitting, $\Delta_{S-O}$, 
of the $p^{3/2}$ and $p^{1/2}$ orbitals. As a result, the SO band is moved towards lower energy with the $\Delta_{S-O}$ values 
(see Table~\ref{table:tab1}) strongly increasing with the atomic number of $X$ atom. Starting from $\Delta_{S-O} = 36$ meV for in 
Mg$_2$Si, through $\Delta_{S-O} \simeq 208$ meV in Mg$_2$Ge, the largest $\Delta_{S-O} \simeq 525$ meV is reached in the heaviest Mg$_2$Sn 
compound. Overall, a good agreement between the KKR results and the experimental data derived from infrared spectroscopy measurements 
is found. 

The second modification of the band structure of Mg$_2X$ induced by the S-O interaction, is removing of HH and LH band degeneracy, which appears 
in some BZ parts (e.g. along $\Gamma$-L-X directions) from the semi-relativistic KKR calculations. This splitting is smaller, 
comparing to the splitting of $p^{3/2}$ and $p^{1/2}$ orbitals at $\Gamma$, however at the L point it can also reach 
considerable values, i.e. 287~meV ($X=$~Sn), 118~meV ($X=$~Ge) and 23~meV ($X=$~Si).

The third and also the most subtle effect, visible actually on an enlarged scale only (Fig.~\ref{fig:Bands4}), is related to 
the modification of the bands shapes, which at first glance suggests a decrease of $m^*$ (as discussed below). On the whole, the three aforementioned 
effects are expected to markedly influence thermopower behaviors in Mg$_2X$ compounds 
(see Sec.\ref{sec:thermo}).

The magnitude of overall relativistic effects as well as the S-O splitting of the band structure near the $\Gamma$ point is well 
illustrated in Fig.~\ref{fig:FS}, where Fermi surfaces (FS) for Mg$_2$Sn at a hole concentration $p = 10^{21}$~cm$^{-3}$ are 
plotted. Similar comparison of FR and SR results for $p-$type Mg$_2$Si and also Mg$_2$Ge are less pronounced 
on corresponding FS and they are not presented. We see that the high velocity (red color in the Fig.~\ref{fig:FS}) hole pocket 
centered at $\Gamma$, becomes the SO band pocket and is shifted out the Fermi surface at the hole concentration 
$p=10^{21}$~cm$^{-3}$. Also, the change in the FS curvature can be easily noticed.

\begin{figure}[htb]
\centering
\includegraphics[width=0.44\textwidth]{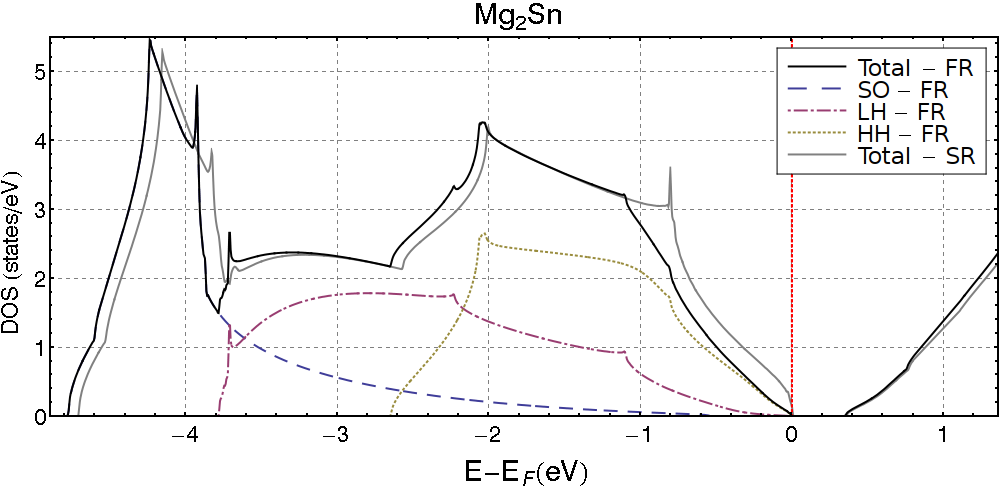}
\vspace{3mm}
\includegraphics[width=0.44\textwidth]{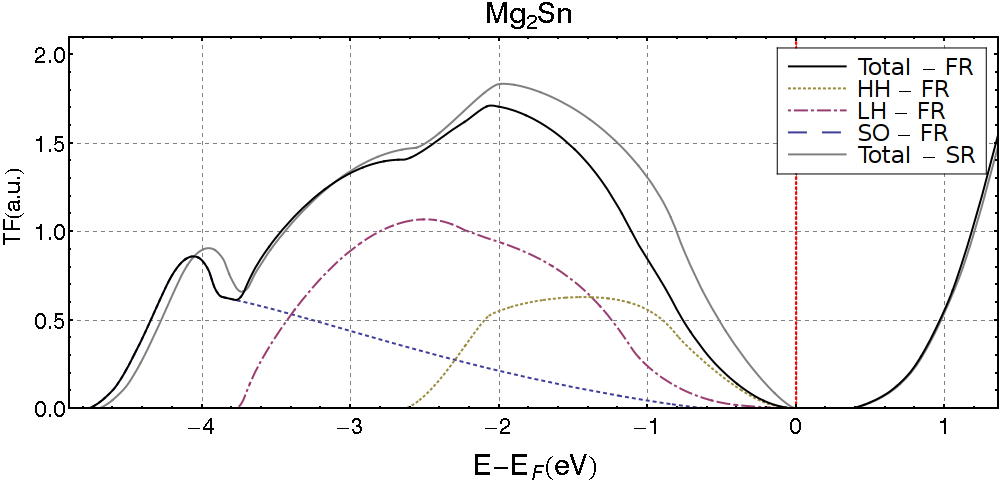}
\caption{(Color online) Density of states (DOS) and transport function (TF) of Mg$_2$Sn with decomposition on bands calculated with full-relativistic (FR) method. 
For comparison DOS and TF are also shown for semi-relativistic (SR) case.}
\label{fig:FTSn}
\end{figure}

The S-O interaction effects, as described above, first of all lower DOS near the valence band edge. This is accompanied by a 
decrease of the DOS effective mass, which likely reduces the thermopower. The effective mass calculated with use of 
Eq.~\ref{eq:mass}, is shown in Fig.~\ref{fig:meff}. The reduction of $m^*$ by the S-O interaction is clearly seen for the Mg$_2$Ge 
and Mg$_2$Sn compounds, where it reaches around 50\% (dropping from $\sim1.1$ to $\sim0.5$ $m_e$ in the latter). 
Such an important reduction can not be explained by the shift of the high velocity SO band, and the change of mass is mainly 
attributed to modification of the $\mathscr{E}(\bf k)$ slope of the HH and LH bands (see, Fig.~\ref{fig:Bands4}). 
The bottom-right panel in Fig.~\ref{fig:meff} shows the effective mass of $p-$type Mg$_2$Sn as resulted from the FR and SR 
calculations, decomposed into all three band contributions. This clearly evidences, that actually the whole contribution to the effective 
mass comes from the HH band. Bearing in mind that effective mass is a additive quantity due to 
formula $m^*_{all}=\left(\sum_i \left. m^*_i\right.^{3/2}\right)^{2/3}$, the modification of bands curvature by the S-O 
interaction is the most important effect for the decrease of $m^*$, more significant than the splitting of bands near
$\Gamma$. The same can be concluded from Fig~\ref{fig:FTSn}, where the reduction of total DOS ($g$) is much larger than the SO 
band contribution, which is very low due to high velocity ($g\sim 1/v$). 

We can also notice an interesting topological feature of the dispersion relations in the FR case: the LH band is convex along  
the $\Gamma$-L direction, close to the valence band edge. In the SR case, the flexion point, at which the $\mathscr{E}(\bf k)$ 
function changes from convex to concave, lies at much deeper energy. The S-O interaction moves this flexion point in the energy 
(and carrier concentration) range important for the TE properties. This convex $\mathscr{E}(\bf k)$ function locally leads 
to the positive effective mass, i.e. there is an electron-like, compensating contribution to the overall hole-like electronic 
properties. This FS feature reduces the LH band effective mass, being another source of lowering of $m^*$. This behaviour resembles 
the case of PbTe-PbS alloy~\cite{pbse-heremans}, where similar topological effect of the local convex shape of the valence 
band leads to the negative Seebeck coefficient at low temperatures.

In Mg$_2$Ge compound the HH band modifications as well as energy splitting of the bands is smaller than in Mg$_2$Sn, and, as a 
consequence, the effective mass reduction is limited to the narrow energy range near the valence band edge.

In Mg$_2$Si the effect is opposite to the previously discussed cases. Fully relativistic treatment yields c.a. $10\%$ rise 
of the value of effective mass, due to the $\mathscr{E}(\bf k)$ slope change, except for the range just below the valence band edge, 
where the SO band moves out and local convex $\mathscr{E}(\bf k)$ function shows up.

The analysis of the effective mass suggests how the spin-orbit interaction will modify the thermopower of the system, 
since $S \propto m^*$. For the Mg$_2$Sn we may expect strong reduction of the $p-$type thermopower, for the Mg$_2$Ge the 
reduction of thermopower is expected only at low concentration in $p-$type materials, and the increase of the Seebeck 
coefficient for Mg$_2$Si can be predicted. Lowering of thermopower in Mg$_2$Sn system can also be deduced from 
transport function in Fig~\ref{fig:FTSn}, where its strong lowering in $p-$type doping is seen in FR calculation. 
It is also worth noting that the SO band has negligible influence on total TF (and therefore thermopower), which is not 
so evident accounting for the fact that TF is a function sensitive to electron's velocity ($\sigma\sim v^2$) and the SO 
band has the highest value of $v$ up to $1.0\times10^6$~ms$^{-1}$ (see next paragraph for detailed calculation of 
thermopower).

As far as the conduction bands are concerned, no significant difference between FR and SR calculations are observed, even in the 
case of Mg$_2$Sn. The Fermi surfaces and also the effective masses do not show any difference for $n-$type doping as large 
as $n=10^{21}$ cm$^{-3}$(see Figs.~\ref{fig:FS}b and \ref{fig:meff}).

\subsection{Thermopower}
\label{sec:thermo}

\begin{table*}[bht]
\caption{
Seebeck coefficient (FR and SR results) at $n=p=10^{20}$cm$^{-3}$ for different temperatures in Mg$_2X$ compounds.}
\centering
\begin{ruledtabular}
\begin{tabular}{p{0.1cm}lccccccccc}
\multicolumn{1}{c}{ }&\multicolumn{1}{c}{ } &\multicolumn{3}{c}{Mg$_2$Si}&\multicolumn{3}{c}{Mg$_2$Ge}&\multicolumn{3}{c}{Mg$_2$Sn}\\\cline{3-5}\cline{6-8}\cline{9-11}
 & & 70 K & 300 K & 900 K & 70 K & 300 K & 900 K & 70 K & 300 K & 900 K \\
\hline
\multirow{3}{*}{ \ \rotatebox[origin=top]{90}{n-type }}& S$_{\text{FullRell}}$ ($\mu$V/K) & -33 & -139 & -291 & -27 & -113 & -273 & -20 & -84 & -214  \\
&S$_{\text{SemiRell}}$  ($\mu$V/K)& -34 & -141 & -290 & -27 & -112 & -269 & -19 & -82 & -207    \\
&S$_{\text{FR}}$/S$_{\text{SR}}$   (-) & 98\% & 99\% & 100\%& 100\% & 101\% & 101\% & 103\% & 102\% & 104\%  \\[0.2ex]
\hline
\multirow{3}{*}{ \ \rotatebox{90}{p-type }} & S$_{\text{FullRell}}$ ($\mu$V/K)& 77 & 208 & 329 & 41 & 154 & 305 & 22 & 94 & 224   \\
&S$_{\text{SemiRell}}$ ($\mu$V/K)& 65 & 191 & 314 & 62 & 189 & 310 & 47 & 156  & 263   \\
&S$_{\text{FR}}$/S$_{\text{SR}}$ (-) & 118\% & 109\% & 105\% & 67\% & 83\% & 98\% & 48\% & 60\% & 85\%  \\
\end{tabular}
\end{ruledtabular}
\label{table:tab2}
\end{table*}

\begin{figure}[htb]
\centering
\includegraphics[width=0.48\textwidth]{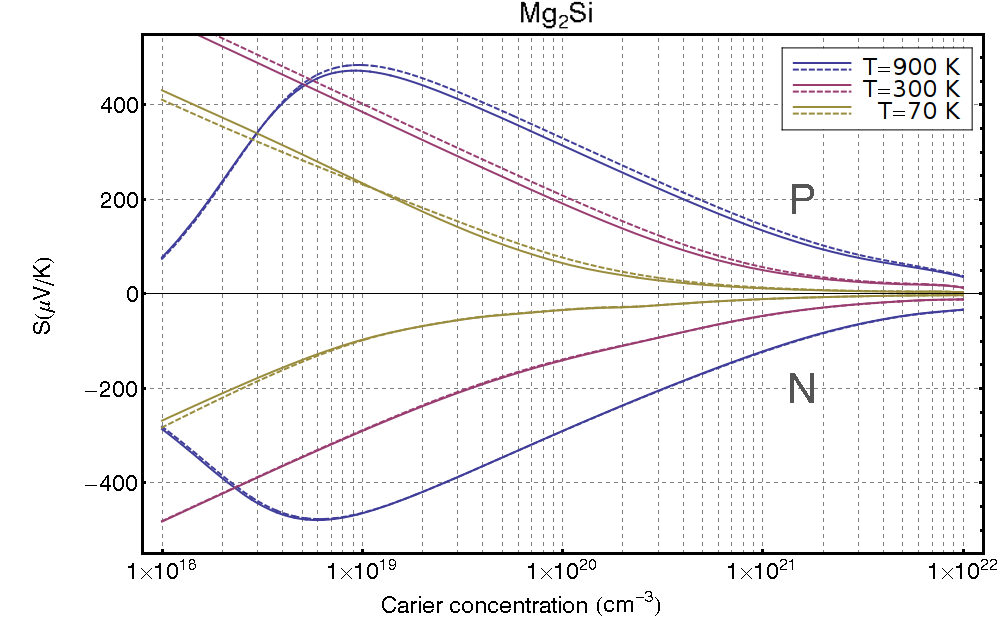}
\includegraphics[width=0.48\textwidth]{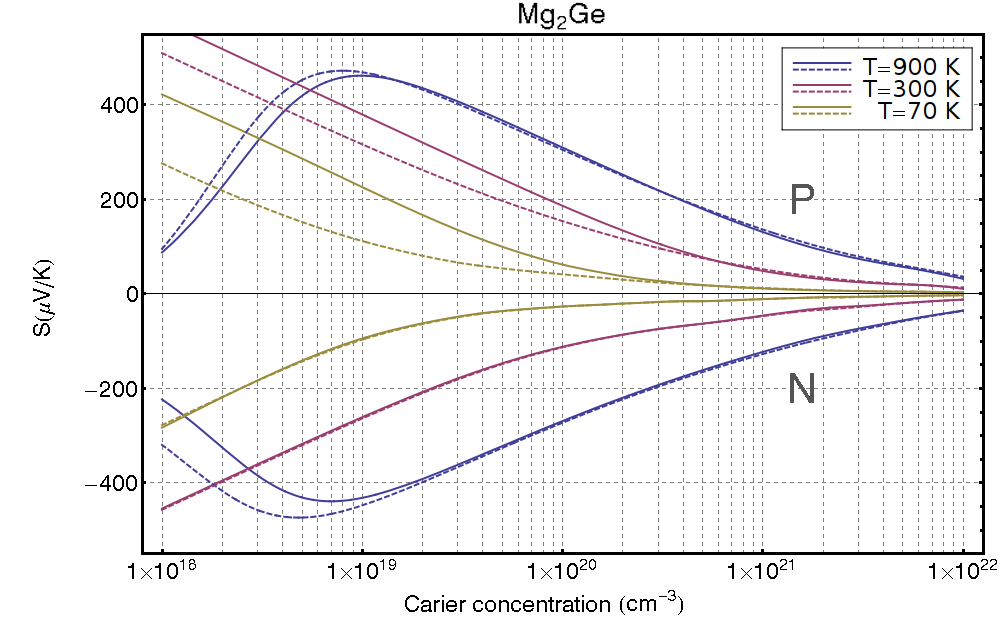}
\includegraphics[width=0.48\textwidth]{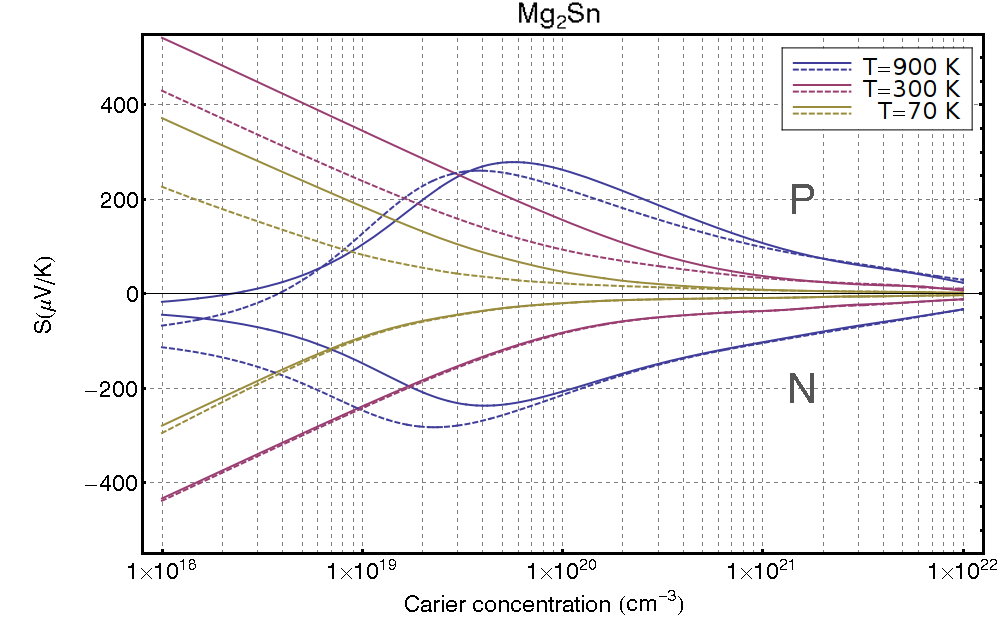}
\caption{(Color online) Thermopower of Mg$_2$Si (top), Mg$_2$Ge (middle) and Mg$_2$Sn (bottom) as a function of carrier concentration for selected temperatures. Solid lines show semi-relativistic and dashed line full-relativistic results, respectively. }
\label{fig:TP1}
\end{figure}

Thermopower of Mg$_2X$ compounds, calculated in the constant relaxation time approximation and using Eqs.~\ref{eq:s}-\ref{eq:sig}, are shown in a wide range of carrier concentration 
for both $n-$ and $p-$type in Fig.~\ref{fig:TP1}. In addition, for the carrier concentrations  $n=p=10^{20}$~cm$^{-3}$ and at three selected temperatures (70~K, 300~K and 900~K) the Seebeck 
coefficient values are gathered in Table~\ref{table:tab2}.
The results of the Boltzmann transport and KKR calculations are consistent with intuitive predictions based on the aforementioned 
analysis of the band splitting and effective mass changes. First of all, for the three Mg$_2X$ compounds there is no significant 
difference between thermopower from FR and SR calculations in the $n-$type materials for high concentration and low temperature. 
The reason is simply the negligible differences between conduction bands induced by the S-O interaction. On the other hand, 
the difference between FR and SR appearing for lower carrier concentration and higher temperature comes from the bipolar effect, which 
accounts for contribution from valence bands, that are apparently different from the FR and SR results. 
The bipolar reduction of the absolute value of $n-$type $S$ is highest for Mg$_2$Sn, where the band gap is the smallest. 
In fully relativistic KKR calculations Seebeck coefficient is less reduced, because of the lower effective mass of holes.

For hole-like thermopower, the splitting of the bands accompanied by the reduction of the effective mass significantly decrease 
the Seebeck coefficient in Mg$_2$Ge and Mg$_2$Sn for $p < 10^{21}$ cm$^{-3}$ in the low and mid-temperature range 
(see also Tab.~\ref{table:tab2}). For Mg$_2$Sn, the S-O effect can lower thermopower even twice, showing that it is 
crucial for the discussion of the TE properties of $p-$type Mg$_2X$ systems. At high temperature, due to the temperature 
blurring of DOS (convoluted with Fermi-Dirac function), the main contribution to thermopower comes from the lower-lying parts 
of dispersion curves, where computed band curvature from FR and SR approaches, are similar. Also, at high temperature the importance 
of the SO band splitting diminishes, which effectively lowers the difference between SR and FR computations. Similar trend is apparently 
observed when the carrier concentration increases and the Fermi level moves deeper into the valence bands and almost no difference 
between SR and FR results is observed for $p > 10^{21}$ cm$^{-3}$.

Interestingly, FR KKR calculations show that in $p-$type Mg$_2$Si both $m^*$ and $S$ values are in principle higher than the values 
from the SR computations. The slightly reduced $S$ is seen only in very low concentration and temperature, where the small area of FS 
near $\Gamma$ point are taken into consideration in electron transport. 
In this case, i.e. $p<1\times10^{19}$, the band curvature modification as well as the S-O splitting, seen in the Fig.~\ref{fig:Bands4}, 
leads to the decrease of $m^*$, and therefore to small decrease of $S$ (see Fig~\ref{fig:TP1} at $T=70$~K and small hole concentration $p$). 
For $p>1\times10^{19}$, these effects are no longer important. At higher temperature the S-O splitting is too small and additionally 
smeared by temperature, i.e. $\Delta_{S-O} = 30$~meV $=k_BT$ for $T\sim350$~K, that it does not alter results from transport function 
integration (Eq.~\ref{eq:s}).


\section{Summary}
\label{sec:Conclusions}
The results of the electronic band structure calculations with the use of fully relativistic (FR) vs. semi-relativistic (SR) full 
potential KKR method for the Mg$_2X$ ($X=$~Si, Ge and Sn) compounds were reported. It was found that the S-O interaction 
splits the valence band structure, with the S-O splitting at $\Gamma$ point, namely 36 meV, 208 meV and 525 meV, well 
corresponding to experimental data\cite{vazquez1968,Lott1965} 30 meV, 200 meV and 480 meV (or 600 meV) in Mg$_2$Si, Mg$_2$Ge and 
Mg$_2$Sn, respectively. The S-O splitting itself, as well as modification in dispersion relation (even more important) of top valence bands, 
i.e. heavy and light hole bands, significantly decreased the effective mass and the Seebeck coefficient. 
In Mg$_2$Ge and Mg$_2$Sn, the analysis of the energy dependent DOS effective mass shows that the S-O interaction 
lowers $m^*$ of holes, and the effect is larger near the valence band edge and for the heaviest compound. 
In general, this leads to the reduction of the $p-$type thermopower, mostly at lower concentration and low or medium range of 
temperature. The thermopower decrease becomes more serious with increasing atomic number of $X$ element, reaching the magnitude 
of $50\%$ in Mg$_2$Sn. On the whole, our KKR calculations combined with Boltzmann transport approach clearly show that the 
relativistic effects are detrimental for thermoelectric performance in $p-$doped Mg$_2$Ge and Mg$_2$Sn. In Mg$_2$Si, the S-O 
interaction slightly increases the effective hole mass, except for the low carrier concentration range. 
This behavior leads to the small increase of thermopower (up to 10\%) for $p > 10^{19}$~cm$^{-3}$. 

In turn, the S-O coupling effect on the conduction bands is negligible in all Mg$_2X$ compounds. Surprisingly, the $n-$type 
thermopower at lower carrier concentrations benefits from the degradation of the $p-$type $S$ due to the reduction of 
the bipolar effect, which is well seen in Mg$_2$Ge and Mg$_2$Sn compounds.

In summary, the full relativistic KKR results revealed that the spin-orbit interaction is significant factor decreasing the 
thermoelectric performance of $p-$type Mg$_2X$ (especially Mg$_2$Sn). They also enlightened the reason why in Mg$_2$(Si-Ge-Sn) 
system the measured $zT$ in $p-$doped samples was much lower ($zT <0.4$)\cite{mouko11} than the values gained in $n-$doped 
materials ($zT \sim 1.4$)\cite{khan13,liu13}.


\begin{acknowledgments}
This work was supported by the Polish National Science Center (NCN) under the grants DEC-2011/02/A/ST3/00124 and UMO-2011/03/N/ST3/02644, 
Thermomag project (FP7-NMP4-SL-2011-263207) co-funded by the European Space Agency and by individual partner organizations as well as by 
the Polish Ministry of Science and Higher Education.
\end{acknowledgments}



\end{document}